\documentclass[epj]{svjour}
\usepackage{graphics}
\def\PRL{Phys. Rev. Lett.}
\def\PRC{Phys. Rev. C}
\def\PRD{Phys. Rev. D}
\def\PLB{Phys. Lett. B}
\def\NPA{Nucl. Phys. A}
\def\NPB{Nucl. Phys. B}
\def\JPG{J. Phys. G: Nucl. Part. Phys.}
\def\IJMPE{Int. J. Mod. Phys. E}
\def\EPJA{Eur. Phys. J. A}
\def\AP{Ann. Phys. (N.Y.)}
\def\AR{Annu. Rev. Nucl. Part. Sci.}
\def\Journal#1#2#3#4{#1 {\bf #2}, #3 (#4)}

\newcommand{\ba}{\begin{eqnarray}}
\newcommand{\ea}{\end{eqnarray}}

\begin{document}
\title{Flavor content of nucleon form factors in a VMD approach}
\author{R. Bijker}  
\institute{Instituto de Ciencias Nucleares, 
Universidad Nacional Aut\'onoma de M\'exico, 
AP 70-543, 04510 M\'exico DF, M\'exico}
\date{Received: date / Revised version: date}
%
\abstract{
The strange form factors of the nucleon are studied in a two-component 
model consisting of a three-quark intrinsic structure surrounded by a 
meson cloud. A comparison with the available experimental world data 
from the SAMPLE, PVA4, HAPPEX and G0 collaborations shows a good overall 
agreement. It is shown that the strangeness contribution to the electric 
and magnetic form factors is of the order of a few percent. 
In particular, the strange quark contribution to the charge radius is small 
$\left< r^2_s \right>_E = 0.005$ fm$^2$ and to the magnetic moment it is  
positive $\mu_s=0.315$ $\mu_N$. 
}
\maketitle

\section{Introduction}

The flavor content of the electromagnetic form factors of the nucleon 
can be studied by combining the nucleon's response to the electromagnetic 
and weak neutral vector currents \cite{Manohar}. In recent experiments, 
parity-violating elastic electron-proton scattering has been used to probe 
the contribution of strange quarks to the structure of the nucleon 
\cite{beckexp,Beise}.  
The strange quark content of the form factors can be determined 
assuming charge symmetry and combining parity-violating asymmetries 
with measurements of the electric and magnetic form factors of the 
proton and neutron. The study of the strange quark content is of special 
interest because it is exclusively part of the quark-antiquark sea. 

The experimental results from the SAMPLE, PVA4, HAPPEX and G0 collaborations 
have shown evidence for a nonvanishing strange quark contribution to the 
structure of the nucleon. In particular, evidence was found that the 
strange magnetic moment of the proton is 
positive \cite{Aniol05b}, suggesting that the strange quarks reduce the 
proton's magnetic moment. This is an unexpected and surprising finding, 
since a majority of theoretical studies favors a negative value \cite{beckth}. 

The aim of this contribution is to study the flavor content of nucleon 
form factors in a VMD approach in which the two-component model of  
electromagnetic nucleon form factors of \cite{BI} is extended to the 
strange sector. The strangeness content is determined via the coupling 
of the strange current to the $\phi$ and $\omega$ mesons \cite{Jaffe}. 
A comparison with the available experimental world data shows a good 
overall agreement for $0 < Q^2 < 1$ (GeV/c)$^2$. 

\section{Nucleon form factors}

Electromagnetic and weak form factors contain the information about the 
distribution of electric charge and magnetization inside the nucleon.  
These form factors arise from matrix elements of the corresponding 
vector current operators
\ba
\left< N \left| V_{\mu} \right| N \right> 
= \bar{u}_N \left[ F_1(Q^2) \, \gamma_{\mu} 
+ F_2(Q^2) \, \frac{i \sigma_{\mu\nu} q^{\nu}}{2M_N} \right] u_N .
\ea
Here $F_{1}$ and $F_2$ are the Dirac and Pauli form factors 
which are functions of the squared momentum transfer $Q^2=-q^2$. 
The electric and magnetic form factors, $G_{E}$ and $G_{M}$, are 
obtained from $F_{1}$ and $F_{2}$ by the relations $G_E=F_1-\tau F_2$ 
and $G_M=F_1 + F_2$ with $\tau=Q^2/4 M_N^2$. 

The Dirac and Pauli form factors are parametrized according to a  
two-component model of the nucleon \cite{BI} in which the external photon 
couples both to an intrinsic three-quark structure described by the form 
factor $g(Q^2)$ and to a meson cloud through the intermediate vector mesons  
$\rho$, $\omega$ and $\phi$. In the original version of the two-component model 
\cite{IJL}, the Dirac form factor was attributed to both the intrinsic structure and 
the meson cloud, and the Pauli form factor entirely to the meson cloud. 
In a modified version \cite{BI}, it was shown that the addition of an intrinsic 
part to the isovector Pauli form factor as suggested by studies of 
relativistic constituent quark models in the light-front approach 
\cite{frank}, improves the results for the elecromagnetic form factors 
of the neutron considerably. 

In order to incorporate the contribution of the isocalar ($\omega$ and 
$\phi$) and isovector ($\rho$) vector mesons, it is convenient to 
first introduce the isoscalar and isovector current operators 
\ba
V_{\mu}^{I=0} &=& \frac{1}{6} \left( \bar{u} \gamma_{\mu} u 
+ \bar{d} \gamma_{\mu} d -2 \bar{s} \gamma_{\mu} s \right) ,
\nonumber\\ 
V_{\mu}^{I=I} &=& \frac{1}{2} \left( \bar{u} \gamma_{\mu} u 
- \bar{d} \gamma_{\mu} d \right) . 
\ea
The corresponding isoscalar Dirac and Pauli form factors depend on 
the couplings to the $\omega$ and $\phi$ mesons 
\ba
F_{1}^{I=0}(Q^{2}) &=& \frac{1}{2} g(Q^{2}) \left[ 
1-\beta_{\omega}-\beta_{\phi} \right.
\nonumber\\
&& \hspace{0.5cm} \left. 
+\beta_{\omega} \frac{m_{\omega }^{2}}{m_{\omega }^{2}+Q^{2}} 
+\beta_{\phi} \frac{m_{\phi}^{2}}{m_{\phi }^{2}+Q^{2}}\right] , 
\\
F_{2}^{I=0}(Q^{2}) &=& \frac{1}{2}g(Q^{2})\left[ 
\alpha_{\omega} \frac{m_{\omega }^{2}}{m_{\omega }^{2}+Q^{2}} 
+ \alpha_{\phi} \frac{m_{\phi}^{2}}{m_{\phi}^{2}+Q^{2}}\right] ,
\nonumber
\label{ff}
\ea
and the isovector ones on the coupling to the $\rho$ meson \cite{BI}
\ba 
F_{1}^{I=1}(Q^{2}) &=& \frac{1}{2}g(Q^{2})\left[ 1-\beta_{\rho} 
+\beta_{\rho} \frac{m_{\rho}^{2}}{m_{\rho}^{2}+Q^{2}} \right] , 
\nonumber\\ 
F_{2}^{I=1}(Q^{2}) &=& \frac{1}{2}g(Q^{2})\left[ 
\frac{\mu_{p}-\mu_{n}-1-\alpha_{\rho}}{1+\gamma Q^{2}} \right.
\nonumber\\
&& \hspace{2.5cm} \left. 
+\alpha_{\rho} \frac{m_{\rho }^{2}}{m_{\rho}^{2}+Q^{2}} \right] .
\ea
The proton and neutron form factors correspond to the sum and 
difference of the isoscalar and isovector contributions, 
$F_i^p=F_i^{I=0}+F_i^{I=i}$ and $F_i^n=F_i^{I=0}-F_i^{I=i}$, respectively. 
This parametrization ensures that the three-quark contribution to the 
anomalous magnetic moment is purely isovector, as given by $SU(6)$. 
The intrinsic form factor is a dipole $g(Q^{2})=(1+\gamma Q^{2})^{-2}$ 
which coincides with the form used in an algebraic treatment of the 
intrinsic three-quark structure \cite{bijker}. 
The large width of the $\rho$ meson which is crucial for the small $Q^{2}$ 
behavior of the form factors, is taken into account in the same way as in 
\cite{BI,IJL}. For small values of $Q^2$ the form factors are dominated by the 
meson dynamics, whereas for large values they satisfy the asymptotic behavior 
of p-QCD, $F_1 \sim 1/Q^4$ and $F_2 \sim 1/Q^6$ \cite{pQCD}.

\section{Flavor content}

The strange quark content of the nucleon form factors arises through 
the coupling of the strange current 
\ba 
V_{\mu}^s = \bar{s} \gamma_{\mu} s ,
\ea 
to the intermediate isocalar vector mesons  $\omega$ and $\phi$ (using the 
convention of Jaffe \cite{Jaffe}). The wave functions of the $\omega$ 
and $\phi$ mesons are given by
\ba
\left| \omega \right> &=& \cos \epsilon \left| \omega_0 \right> 
- \sin \epsilon \left| \phi_0 \right> ,
\nonumber\\
\left| \phi \right> &=& \sin \epsilon \left| \omega_0 \right> 
+ \cos \epsilon \left| \phi_0 \right> ,
\ea
where the mixing angle $\epsilon$ represents the deviation from the 
ideally mixed states  
$\left| \omega_0 \right>=\left( u \bar{u} + d \bar{d} \right)/\sqrt{2}$ 
and $\left| \phi_0 \right> = s \bar{s}$. 
Under the assumption that the strange form factors have the same form as 
the isoscalar ones, the Dirac and Pauli form factors that correspond to the 
strange current are expressed as the product  
of an intrinsic part $g(Q^2)$ and a contribution from the vector mesons 
\ba
F_{1}^{s}(Q^{2}) &=& \frac{1}{2}g(Q^{2})\left[ 
\beta_{\omega}^s \frac{m_{\omega}^{2}}{m_{\omega }^{2}+Q^{2}} 
+\beta_{\phi}^s \frac{m_{\phi}^{2}}{m_{\phi }^{2}+Q^{2}}\right] , 
\nonumber\\
F_{2}^{s}(Q^{2}) &=& \frac{1}{2}g(Q^{2})\left[ 
\alpha_{\omega}^s \frac{m_{\omega}^{2}}{m_{\omega }^{2}+Q^{2}}
+\alpha_{\phi}^s \frac{m_{\phi}^{2}}{m_{\phi }^{2}+Q^{2}}\right] .
\label{sff}
\ea
The isocalar and strange couplings appearing in Eqs.~(\ref{ff}) and (\ref{sff}) 
are not independent of one another, but depend on the same nucleon-meson and 
current-meson couplings \cite{Jaffe}. In addition, they are constrained 
by the electric charges and magnetic moments of the nucleon which leads 
to two independent isoscalar couplings
\ba
\alpha_{\omega} &=& \mu_p + \mu_n -1 - \alpha_{\phi} ,
\nonumber\\
\beta_{\phi} &=& - \beta_{\omega} \tan \epsilon/\tan(\theta_0+\epsilon) , 
\label{coef1}
\ea
from which the strange couplings can be obtained as \cite{Jaffe} 
\ba
\beta_{\omega}^s/\beta_{\omega} =  
\alpha_{\omega}^s/\alpha_{\omega} &=& 
-\sqrt{6} \, \sin \epsilon/\sin(\theta_0+\epsilon) ,
\nonumber\\
\beta_{\phi}^s/\beta_{\phi} =  
\alpha_{\phi}^s/\alpha_{\phi} &=& 
-\sqrt{6} \, \cos \epsilon/\cos(\theta_0+\epsilon) .  
\label{coef2}
\ea
with $\tan \theta_0 = 1/\sqrt{2}$. The mixing angle $\epsilon$ can be determined either 
from the radiative decays of the $\omega$ and $\phi$ mesons \cite{Jain,Iachello,Harada} 
or from their strong decays \cite{Gobbi}. The value used here is $\epsilon=0.053$ 
rad (3.0$^{\circ}$) \cite{Jain}. 

Finally, the contributions of the up and down quarks to the electromagnetic 
form factors can be obtained from 
\ba
G^u_{E/M} &=& 2G^p_{E/M} + G^n_{E/M} + G^s_{E/M} ,
\nonumber\\
G^d_{E/M} &=& G^p_{E/M} + 2G^n_{E/M} + G^s_{E/M} .
\ea

\section{Results}

In order to calculate the nucleon form factors in the two-component model the 
five coefficients, $\gamma$ from the intrinsic form factor, $\beta_{\omega}$ and 
$\alpha_{\phi}$ from the isoscalar couplings, and $\beta_{\rho}$ and 
$\alpha_{\rho}$ from the isovector couplings, are determined in a least-square 
fit to the electric and magnetic form factors of the proton and the neutron using the 
same data set as in \cite{BI}. The electromagnetic form factor of the proton 
and neutron are found to be in good agreement with experimental data \cite{iguazu}.  
According to Eq.~(\ref{coef2}), the strange couplings can be determined from the 
fitted values of the isoscalar couplings to be 
$\beta_{\phi}^s=-\beta_{\omega}^s=0.202$,  $\alpha_{\phi}^s=0.648$ and 
$\alpha_{\omega}^s=-0.018$ \cite{iguazu,jpg}. 

\begin{figure}
\resizebox{0.5\textwidth}{!}{%
\includegraphics{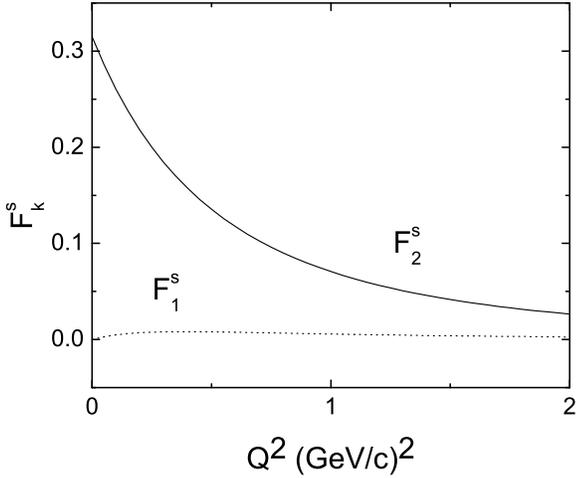}}
\caption{Strange Dirac and Pauli form factors, $F_1^s$ (dotted line) and 
$F_2^s$ (solid line).}
\label{f12s}
\end{figure}

Figure~\ref{f12s} shows the strange Dirac and Pauli form factors as a function 
of the momentum transfer $Q^2$.  
Whereas the Pauli form factor is dominated by the coupling to the $\phi$ 
meson ($\alpha_{\phi}^s \gg \alpha_{\omega}^s$), the Dirac form factor is 
small due to a cancelation between the contributions from the $\omega$ and $\phi$ 
mesons ($\beta_{\phi}^s=-\beta_{\omega}^s$). The qualitative features of these 
form factors can be understood in the limit of ideally 
mixed mesons, {\em i.e.} zero mixing angle $\epsilon=0$ (in comparison to 
the value of $\epsilon=3.0^{\circ}$ used in Figure~\ref{f12s}). Since in this case  
$\beta_{\phi}^s=\beta_{\omega}^s=\alpha_{\omega}^s=0$, the Dirac form factor 
vanishes identically and the Pauli form factor depends only on the tensor coupling 
to the $\phi$ meson, $\alpha_{\phi}^s$.   

The behavior of $F_1^s$ and $F_2^s$ in Figure~\ref{f12s} is quite different from 
that obtained in other theoretical approaches, especially for the strange Pauli 
form factor. Almost all calculations give negative values for $F_2^s$ for the same 
range of $Q^2$ values \cite{Jaffe,Park,Garvey,Forkel,Hammer,Lyubovitskij},  
with the exception of the meson-exchange model \cite{Meissner} 
and the $SU(3)$ chiral quark-soliton model \cite{Silva}. 
In the former case, the values of $F_2^s$ are about two orders of 
magnitude smaller than the present ones, whereas in the latter $F_2^s$  
is positive for small values of $Q^2$, but changes sign around $Q^2=0.1-0.3$ 
(GeV/c)$^2$.

\begin{figure}
\resizebox{0.5\textwidth}{!}{%
\includegraphics{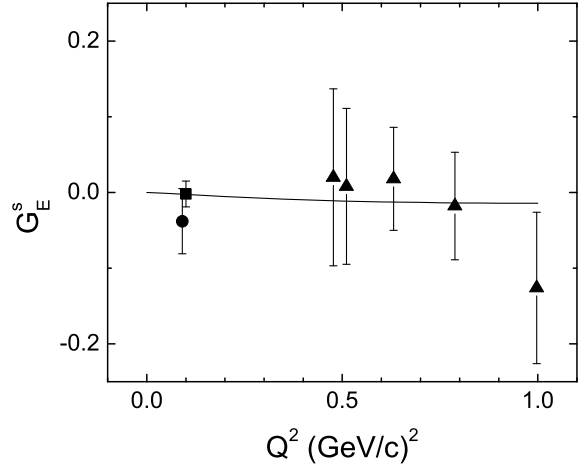}} 
\caption{Comparison between theoretical and experimental values of the strange 
electric form factor. The experimental values are taken from \cite{Aniol05a} 
(circle), \cite{Frascati} (triangle) and \cite{Paschke} (square).}
\label{GEs}
\end{figure}

\begin{figure}
\resizebox{0.5\textwidth}{!}{%
\includegraphics{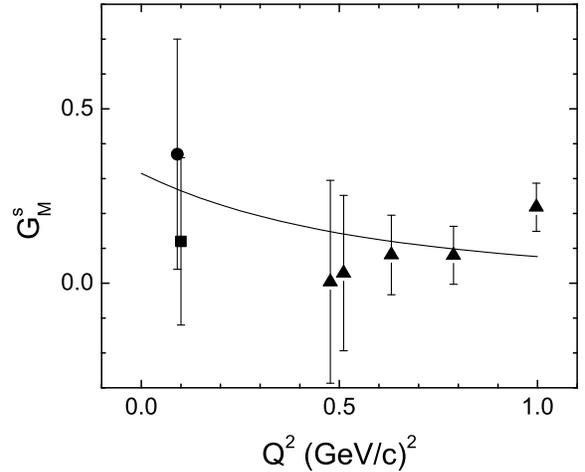}}
\caption{Comparison between theoretical and experimental values of the strange 
magnetic form factor. The experimental values are taken from \cite{Spayde} 
(circle), \cite{Frascati} (triangle) and \cite{Paschke} (square).}
\label{GMs}
\end{figure}

Figures~\ref{GEs} and \ref{GMs} show the strange electric and magnetic form 
factors as a function of $Q^2$. The theoretical values for $G_E^s$ are small 
and negative, in agreement with the experimental results of the HAPPEX  
Collaboration in which $G_E^s$ was determined in parity-violating 
electron scattering from $^{4}$He. The experimental values, 
$G_E^s=-0.038 \pm 0.042 \pm 0.010$ measured at $Q^2=0.091$ (GeV/c)$^2$ 
\cite{Aniol05a} and, more recently,  
$G_E^s=-0.002 \pm 0.017$ at $Q^2=0.1$ (GeV/c)$^2$ \cite{Paschke} 
are consistent with zero. 
 
The values of $G_M^s$ are positive, since they dominated by the 
contribution from the Pauli form factor. Experimental evidence 
from the SAMPLE and HAPPEX collaborations gives a positive value of 
the strange magnetic form factor at $Q^2=0.1$ (GeV/c)$^2$ of  
$G_M^s=0.37 \pm 0.20 \pm 0.26 \pm 0.07$ \cite{Spayde} and 
$G_M^s=0.12 \pm 0.24$ \cite{Paschke}, respectively. 
The other experimental values of $G_E^s$ and $G_M^s$ in Figs.~\ref{GEs} 
and \ref{GMs} for $0.4 < Q^2 < 1.0$ (GeV/c)$^2$ were obtained 
\cite{Frascati,Pate} by combining the (anti)neutrino data from E734 
\cite{Ahrens} with the parity-violating asymmetries from HAPPEX \cite{Aniol04} 
and G0 \cite{Armstrong}. 
The theoretical values are in good overall agreement with the 
experimental ones for the entire range $0 < Q^2 < 1$ (GeV/c)$^2$. 

\begin{table}
\caption{Comparison between theoretical and experimental values of 
strange form factors $G_E^s + \eta G_M^s$.}
\label{HappexA4}
\begin{tabular}{ccccc}
\hline\noalign{\smallskip}
$Q^2$ & $\eta$ & \multicolumn{2}{c}{$G_E^s + \eta G_M^s$} & \\
(GeV/c)$^2$ & & Present & Experiment & Reference \\
\noalign{\smallskip}\hline\noalign{\smallskip}
0.099 & 0.080 & 0.019 & $0.030 \pm 0.028$ & \cite{Aniol05b} \\
0.108 & 0.106 & 0.025 & $0.071 \pm 0.036$ & \cite{Maas2} \\
0.230 & 0.225 & 0.042 & $0.039 \pm 0.034$ & \cite{Maas1} \\
0.477 & 0.392 & 0.047 & $0.014 \pm 0.022$ & \cite{Aniol04} \\
\noalign{\smallskip}\hline
\end{tabular}
\end{table}

\begin{figure}
\resizebox{0.5\textwidth}{!}{%
\includegraphics{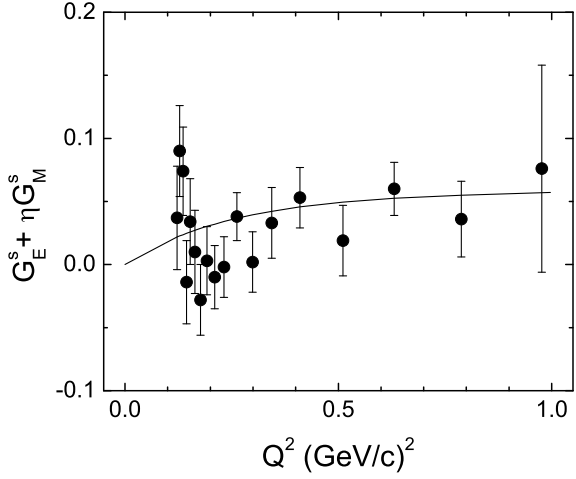}} 
\caption{Comparison between theoretical and experimental values of 
strange form factors $G_E^s + \eta G_M^s$.  
The experimental values were measured by the G0 Collaboration 
\cite{Armstrong}.}
\label{G0}
\end{figure}

Table~\ref{HappexA4} and Figure~\ref{G0} show the results obtained 
by the PVA4, HAPPEX and G0 collaborations for a linear combination  
of the strange electric and magnetic form factors $G_E^s+\eta G_M^s$. 
Also in this case, there is a good agreement between the 
calculated values and the experimental data.

In the majority of theoretical analyses, the strangeness contribution to the 
nucleon is discussed in terms of the static properties, the strange magnetic 
moment $\mu_s$ and the strangeness radius $\left< r_s^2 \right>$. 
Most theoretical studies agree on a small negative strangeness radius and a 
moderate negative strange magnetic moment \cite{beckth}, 
whereas the results of a combined fit of the strange electric and magnetic 
form factors measured by SAMPLE, PVA4 and HAPPEX at $Q^2 \sim 0.1$ (GeV/c)$^2$, 
$G_M^s(0.1)=0.55 \pm 0.28$ and $G_E^s(0.1)=-0.01 \pm 0.03$ \cite{Aniol05b},  
indicate the opposite sign for both $\mu_s$ and $\left< r_s^2 \right>$. 
Recent lattice calculations give small negative values of the strange 
magnetic moment $\mu_s=G_M^s(0)= -0.046 \pm 0.019$ $\mu_N$ \cite{Leinweber1} and 
the strange electric form factor $G_E^s(0.1)=-0.009 \pm 0.028$ \cite{Leinweber2}. 

\begin{figure}
\resizebox{0.5\textwidth}{!}{%
\includegraphics{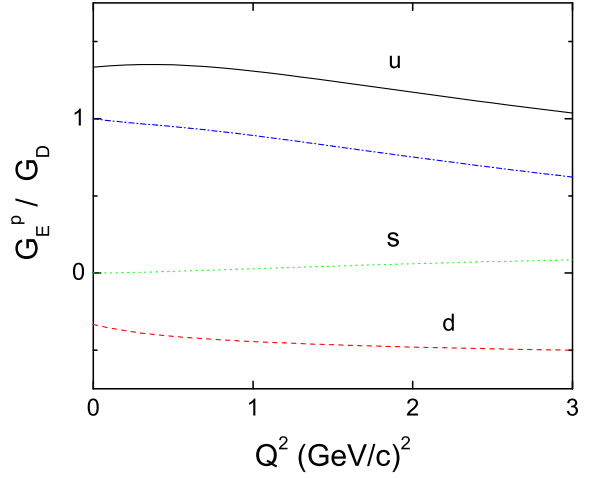}} 
\caption{Flavor decomposition of the proton electric form factor
$G_E^p/G_D$ with $G_D=1/(1+Q^2/0.71)^2$. }
\label{flavorgep}
\end{figure}

\begin{figure}
\resizebox{0.5\textwidth}{!}{%
\includegraphics{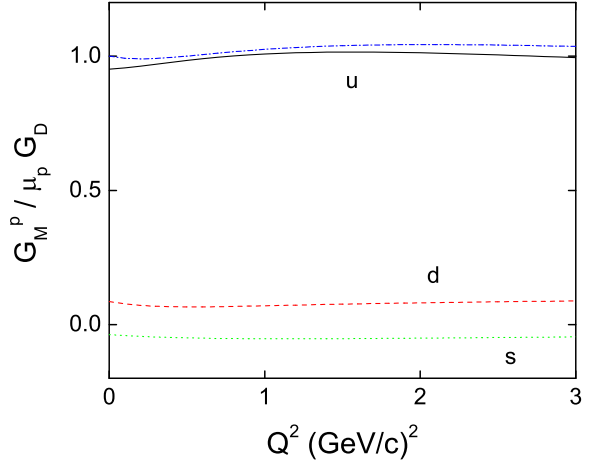}} 
\caption{Flavor decomposition of the proton magnetic form factor
$G_M^p/\mu_p G_D$ with $G_D=1/(1+Q^2/0.71)^2$.}
\label{flavorgmp}
\end{figure}

Figures~\ref{flavorgep} and \ref{flavorgmp} show the flavor decomposition  
of the electric and magnetic form factors of the proton. Note, that in comparison 
with Figs.~\ref{GEs} and \ref{GMs} the flavor form factors have been multiplied 
by the quark electric charges, so that their sum gives the total form factor. 
The contribution of the strange quarks to the proton form factors is small 
for the entire range of $Q^2$ values and of the order of a few percent of the total. 
 
In the present approach, the strangeness contribution to the magnetic moment 
and the charge and magnetic raddi is given by \cite{jpg}
\ba
\mu_s &=& \frac{1}{2} (\alpha_{\omega}^s+\alpha_{\phi}^{s}) 
= 0.315 \, \mu_N , 
\nonumber\\
\left< r^2_s \right>_E   
&=& 3 \beta_{\phi}^{s} \left( \frac{1}{m_{\phi}^2}-\frac{1}{m_{\omega}^2} \right) 
+\frac{3}{4M_N^2}(\alpha_{\omega}^{s}+\alpha_{\phi}^{s})
\nonumber\\ 
&=& 0.005 \mbox{ fm}^2 ,
\nonumber\\
\left< r^2_s \right>_M &=&  
6 \left[ 2\gamma + \frac{\beta_{\phi}^{s}+\alpha_{\phi}^{s}}
{\alpha_{\omega}^s+\alpha_{\phi}^{s}} \frac{1}{m_{\phi}^2}
+ \frac{\beta_{\omega}^{s}+\alpha_{\omega}^{s}}
{\alpha_{\omega}^s+\alpha_{\phi}^{s}} \frac{1}{m_{\omega}^2} \right] 
\nonumber\\
&=& 0.410 \mbox{ fm}^2 .
\label{strangeness}
\ea
The strange magnetic moment does not depend on the mixing angle $\epsilon$ 
\cite{jpg} and its sign is determined by the sign of the tensor coupling 
$\alpha_{\phi}^s$ ($\gg \alpha_{\omega}^s$). 
The sign of the strangeness contribution 
to the magnetic moment and the charge radius is in agreement with the available 
experimental data. A positive value of the strange magnetic moment seems to 
preclude an interpretation in terms of a $uuds\bar{s}$ fluctuation into a 
$\Lambda K$ configuration \cite{Riska}. On the other hand, an analysis of 
the magnetic moment of $uuds\bar{s}$ pentaquark configurations belonging 
to the antidecuplet gives a positive strangeness contribution for states with 
angular momentum and parity $J^P=1/2^+$, $1/2^-$, and negative for $3/2^+$ 
states \cite{BGS}.  

\section{Summary and conclusions}

In this contribution, the flavor content of nucleon form factors was 
studied in a VMD approach in which the two-component model of Bijker 
and Iachello for the electromagnetic nucleon form factors \cite{BI} 
is combined with the method proposed by Jaffe to determine the strangeness 
content via the coupling of the strange current to the $\phi$ and $\omega$ 
mesons \cite{Jaffe}. The strange couplings are completely fixed by 
the electromagnetic form factors of the proton and neutron. 

The good overall agreement between the theoretical and experimental 
values for the electromagnetic form factors of the nucleon and their 
strange quark content shows that the two-component model provides a 
simultaneous and consistent description of the electromagnetic and 
weak vector form factors of the nucleon. It was shown, that the strangeness 
contribution to the charge and magnetization distributions is of the 
order of a few percent. In particular, the strange magnetic moment 
is found to be positive, in contrast with most theoretical studies, 
but in agreement with the presently available experimental information 
from parity-violating electron scattering experiments. 

Future experiments on parity-violating electron scattering to backward 
angles and neutrino scattering will make it possible to determine the 
contributions of the different quark flavors to the electric, magnetic 
and axial form factors, and thus to provide new insight into the complex 
internal structure of the nucleon. 

\section*{Acknowledgments}
This work was supported in part by a research grant from CONACYT, Mexico.

\end{document}